# Soliton Frequency Combs in Elastomer Membrane-Cavity Optomechanics

Sasan Rahmanian[1], Hamza Mouharrar[1], Amin Alibakhshi[2], Zeeshan Iqbal[1,3], Luis Saucedo-Mora[2,4,5], Francisco Javier Montáns[2,6], and Jan Awrejcewicz [7†]

[1]Department of Systems Design Engineering, University of Waterloo, Waterloo, ON N2L 3G1, Canada

[2]Escuela Técnica Superior de Ingeniería Aeronáutica y del Espacio, Universidad Politécnica de Madrid, Pza.Cardenal Cisneros, 28040 Madrid, Spain

[3]Department of Physics, Government College University, Katchery Road, Lahore 54000, Pakistan

[4]Department of Materials, University of Oxford, Parks Road, Oxford, OX1 3PJ, UK

[5]Department of Nuclear Science and Engineering, Massachusetts Institute of Technology, MA 02139, USA

[6]Department of Mechanical and Aerospace Engineering, Herbert Wertheim College of Engineering, University of Florida, Gainesville, FL, 32611, USA

[7]Department of Automation, Biomechanics and Mechatronics, Lodz University of Technology, 1/15 Stefanowskiego St., 90-924, Lodz, Poland

*Sasan Rahmanian and Hamza Mouharrar have equally contributed to this work.

## Abstract

**Solitons, arising from nonlinear wave-matter interactions, stand out for their intrinsic stability during wave propagation and exceptional spectral characteristics[1]. Their applications span diverse physical systems, including telecommunications[2], atomic clocks[3,4], and precise measurements[5,6]. In recent years, significant strides have been made in developing cavity-optomechanics based approaches to generate optical frequency combs (FCs). In this study, we present an innovative approach, never explored before, that leverages elastomer membrane (EM)-cavity optomechanics to achieve the generation of soliton FCs, a highly sought-after phenomenon in the realm of nonlinear wave-matter interactions. Our method**

**represents a significant breakthrough due to its streamlined simplicity, relying on a single continuous-wave (CW) laser pump and an externally applied acoustic wave exciting an EM-cavity, which gives rise to phonons, quantized vibrational energy states intrinsic to the elastomer's crystalline lattice structure. The mechanical resonator and electromagnetic cavity resonance are parametrically coupled within the microwave frequency range, collectively orchestrate the process of soliton FCs formation with remarkable efficiency. Numerical simulations and experimental observations demonstrate the emergence of multiple stable localized opto-mechanical wave packets, characterized by a narrow pulses time-domain response. Crucially, by setting the acoustic wave frequency to match the natural frequency of the EM resonator, the solitons' teeth are precisely spaced, and the EM's motion is significantly amplified, giving rise to a Kerr medium. The successful realization of optomechanical stable solitons represents a monumental advancement with transformative potential across various fields, including quantum computing and spectroscopy.**

## Introduction

Over the last 25 years, optical frequency combs (FCs) have made a resounding impact, shaping fields from metrology to astronomy[10,15,16]. Their fixed frequency coherence established them as essential tools in labs worldwide. Recently, Kerr optical FCs have harnessed optical solitons to create stable spectral lines, benefiting timing, sensing[17–19], microscopy[20–23], resonance stabilization[9,15,16], spectroscopy[24–26], and quantum information processing[27,28]. Optical solitons, arising in nonlinear media such as optical fibers or waveguides, maintain their integrity through a delicate balance between nonlinearity and dispersion, a key underpinning of their stability and spectral quality.

The creation of Kerr optical solitons in microcavities has ignited interest in chip-scale comb generation, providing a recent and promising avenue for optical FCs creation[27]. Recently, attention has shifted to the exploration of soliton FCs within optomechanical resonators, leveraging their strong nonlinear dynamics[2,5]. In optomechanical-cavities, enhanced radiation pressure force results from coupling the motion of a mechanical resonator to the light field in an optical cavity. This dynamic back-action, where photons interact with mechanical modes, offering prospects for a diverse range of applications: from precision-sensitive measurements[18,19,29] to sophisticated quantum manipulation[28,30], and innovative soliton creation[31]. This paradigm also offers a highly auspicious route for frequency conversion, transitioning between microwaves-to-optical or radio-frequency-to-optical realms[10,11], with applications in communication[7,12] and serving as a bridge for interfacing quantum systems[13,14]. Optomechanical coupling induces mechanical nonlinearity, fostering tailored modal dispersion supporting localized optomechanical wave packets or solitons. Achieving this requires a delicate balance of phonon gain, loss, and high optomechanical nonlinearity[32,33]. Cavity optomechanics, once focused on cooling, now explores a "heating" regime, where red-detuned photons drive phonon emission[2], leading to phenomena like phonon lasing, nonlinear optics, and FCs based on surface acoustic waves[34–36]. More recently, researchers were delving into exceptional materials, such as elastomers, to fully unleash frequency comb potential. These intelligent materials, with responsive deformation, are primed for integration into soft robotics[37], wearable electronics[38,39], serving as versatile sensors, actuators, and energy harvesters[40].

This study addresses the generation of soliton FCs within an elastomer membrane (EM)-cavity optomechanical setup, a novel frontier in scientific inquiry. Utilizing transparent elastomer material VHB-4910, we control light transmissibility through mechanical pre-stretch, creating a balanced optomechanical cavity. Carbon grease on the EM's base captures photons while allowing laser light transmission. A 1 μm thin gold layer on the cavity floor reflects laser light, generating counter-propagating photon pairs that interact with the elastomer's mechanical mode. The EM's motion is amplified by an acoustic wave, an external optical parametric amplifier (OPA), which modulates the optical mode within the microwave frequency range, resulting in optomechanical coupling (Figure 1). Tuning the acoustic excitation frequency precisely to the membrane's natural frequency parametrically modulates the electromagnetic optical mode, a robust optomechanical Kerr medium emerges which results in generating multiple soliton FCs at harmonics of the laser pump frequency. These combs display uniform peak spacing akin to the acoustic wave frequency. This exploration underscores the versatility of elastomer membrane-cavity optomechanics, inspiring fresh perspectives and cross-disciplinary applications, particularly in spectrometers and quantum computing devices. Despite persistent challenges, this work achieves a milestone in creating stable and efficient optomechanical soliton FCs using simple and cost-effective devices.

## Results and Discussion

The device employed in this study is constructed from elastomer, a hyperelastic material (see the method section for detailed information). Accurate measurement of the mechanical transverse vibration resonant modes of the EM resonator is crucial for implementing our proposed technique to generate optomechanical FCs. To accomplish this, an acoustic signal generator is used to externally excite the membrane structure, revealing the first and second mechanical natural frequencies. In Figure 2, we present the time-domain velocity response of the membrane's central point while sweeping the applied acoustic wave frequency from 2.5 kHz to 4.5 kHz, maintaining an acoustic intensity of 105 dB (equivalent to 3 Pa). This confirms the membrane elastomer's flexibility, demonstrating its response to lower pressure and small forces. The frequency-response curve is extracted by evaluating the root mean square (RMS) of the recorded time-history data. Consistent with linear vibration theory, external harmonic forcing at frequency $f$ induces a mechanical structure to exhibit harmonic responses at the same frequency $f$, characterized by varying amplitudes and phases across degrees-of-freedom. Mechanical resonance occurs when the forcing frequency aligns with a natural frequency of the structure. As shown in Figure 2(c), we determine the EM's first natural frequency to be $f_m^1 = 3.5$ kHz. Similarly, by sweeping the acoustic frequency within the range of 8 kHz to 10 kHz, we identify the second natural frequency of the membrane resonator at $f_m^2 = 9$ kHz, as depicted in Figure 2(e). Furthermore, analysing the half-power bandwidth of the frequency-response curves reveals mechanical Q-factors of 10.5 and 21 for the first and second modes, respectively.

Figure 3(a) illustrates the schematic for the experimental technique employed for EM-cavity system to generate multi-soliton optomechanical FCs within its dynamic response. A continuous-wave (CW) laser beam (3.505 MHz, 5mW) interacts with the EM resonator, causing interference phenomena between the incident and reflected laser beams. This interference gives rise to the generation of counter-propagating photon pairs, which actively interact with the motion of the elastomer membrane. To excite the mechanical vibration modes of the EM structure within low frequency range, we apply an external acoustic force that is precisely matched to the EM's mechanical modes natural frequencies, initially targeting the first mode and later the second mode. These mechanical vibrations, in turn, lead to the production of phonons quantized vibration energy states that are inherent within the crystalline lattice of the elastomer. As we gradually increase the amplitude of the external acoustic force (70 dB to 105 dB), the motion of the EM intensifies, resulting in a higher density of phonons. These phonons effectively enhance the mechanical waves within the optomechanical cavity leading to the modulation of the optical resonance through interactions with the counter-propagating photon pairs, effectively forming a nonlinear Kerr medium.

Upon amplification of optomechanical nonlinearity within the cavity medium, a stable and localized optomechanical wave packet emerges, referred to as an optomechanical soliton. However, the acoustic excitation frequency is precisely tuned to modulate and align with the membrane's natural frequency, thereby resulting in the creation of multiple soliton frequency combs at harmonics of the laser pump frequency, displaying uniform peak spacing akin to the acoustic wave frequency. Within the soliton regime, the shape and characteristics of the solitons exhibit significant variations based on the pump power and optical intensity within the cavity. These factors, in turn, have a direct impact on the strength of nonlinearity and mode dispersion. The soliton system's dynamics can be described by the ensuing set of non-dimensional and nonlinearly coupled ordinary differential equations (ODEs) (numerical results are available in supplementary material S1),

$$\frac{d\alpha}{dt} = \left(i\big(\omega_L - \omega(w)\big) - \chi_1\right)\alpha + \chi_2 \tag{1}$$

$$\ddot{\mathbf{p}} + c_u\,\dot{\mathbf{p}} + \lambda^2\mathbf{p} + (\mathbb{K}^u:\mathbf{Q}):\mathbf{Q} = \mathbf{0} \tag{2}$$

$$\ddot{\mathbf{Q}} + c_w\dot{\mathbf{Q}} + \mathbb{M}:\mathbf{Q} + (\mathbb{N}.\,\mathbf{p}):\mathbf{Q} + \big((\mathbb{L}:\mathbf{Q}):\mathbf{Q}\big):\mathbf{Q}$$
$$= \cos(\omega t)\mathbf{f} + \frac{\Gamma}{\big(1 + Rw(0,0;t)\big)^2}|\alpha(t)|^2 \tag{3}$$

where t is the time variable, α denotes the optical modal coordinate, w is the membrane transverse displacement, f is the acoustic force and ω is the optical resonance frequency. $c_u$ and $c_w$ are the viscous damping coefficients along the in-plane radial and out-of-plane transverse motions, respectively. The derivation of the equations of motion and their corresponding parameters and tensors are discussed in detail in the supplementary material.

The frequency spectrum of the EM reveals up to 11 harmonics of the pump frequency, arising from the interaction with reflected optical light, representing higher-order optical modes, Figure 2(b). A deeper understanding of this phenomenon is provided by the optomechanical dynamic equations elaborated in the supplementary material. The coupling between optical power and EM transverse displacement is underpinned by mechanical and Kerr-coupling nonlinearity, represented by nonlinear quadratic terms (depicted in optomechanical equations (ES22) and (ES23) in supplementary material S1).

The optomechanical system operates autonomously, where optical power externally excites the optomechanical cavity system, and mechanical motion reciprocally stimulates the optical facet. This bidirectional coupling establishes the basis for intricate interactions between the mechanical and optical fields. Soliton FCs emerge for sufficiently high pump power and EM vibrations. To amplify the Kerr nonlinear couplings within the optomechanical system, we employed an acoustic wave excitation ($|f_{ac}| = 110$ dB, $f_{ac} = f_m^1 = 3.5$ kHz) that intensified the EM resonator motion and heightened the interaction between photons and mechanical modes. The heightened Kerr nonlinear optomechanical interaction triggers the generation of multiple soliton FCs. These combs align at integer multiples of the laser pump frequency, with frequency spacing precisely mirroring the acoustic wave frequency, Figure 3(c). Figurewise, the solitons' width increases with their center frequency, with power directly tied to membrane velocity amplitude. Solitons at harmonics with more significant optical intensity, like the second, fourth, and eighth solitons, yield more power compared to weaker counterparts.

All generated solitons are phase-coherent and share equal spacing, mirroring the acoustic wave frequency. Figure 3(f) depicts a zoom-in view of the initial 0-150 kHz spectrum interval reflecting the mechanical FCs (time domain analysis is illustrated in Figure ED 2). The primary peak corresponds to the acoustic forcing frequency set at ($f_{ac} = 3.5$ kHz), followed by 35 equidistant higher-harmonic combs arising from membrane large deformation and material/geometrical nonlinearities. These emerge during relatively large-amplitude excitation, showcasing material and geometrical nonlinearities in the EM's dynamics. Focusing on the optomechanical response of the EM,

Figures 3(e) and 3(f) provide insight into the second and fourth solitons. Specifically, the 14.02 MHz (fourth soliton) frequency comb encompasses approximately 260 comb lines, from 13.58 MHz to 14.43 MHz, maintaining an FSR of 3.5 kHz. Shifting to the eighth soliton at 28.04 MHz, it spans an even broader frequency range with around 500 combs, with the same FSR of 3.5 kHz.

By adjusting the frequency of the external acoustic wave to match the elastomer membrane's second resonant mode, the solitons' FSR is extended to FSR $= f_{ac} = f_m^2 = 9$ kHz, Figure 4. The Kerr optomechanical nonlinear interaction leads to the generation of higher harmonics of the laser pump frequency, spanning up to eleven times the fundamental laser frequency. Maintaining an acoustic wave intensity of 105 dB, similar to the excitation technique for the first mechanical mode, the generation of multi-soliton FCs becomes more efficient when exciting the second mechanical mode, which not only enhances the power of the combs but also broadens the frequency range of the soliton combs, leading to a more pronounced and coherent Kerr-comb soliton generation. As before, the spacing of the optomechanical comb lines aligns with the mechanical mode, coinciding with the acoustic wave excitation frequency. Notably, a mechanical comb emerges at lower frequencies, spanning from 0 to 200 kHz. For instance, the eighth soliton with a frequency of 28.04 MHz encompasses approximately 250 comb lines, spanning over 2 MHz with an FSR of 9 kHz. Comparing Figures 3(e), 3(f), 4(e), and 4(f), it's evident that soliton comb power increases, peaking at both ends of the generated solitons, driven by the energy directed to the membrane's second resonant mode. Numerical simulations of the developed mathematical equations are provided in the supplementary material to validate the experimental results.

In Figure 5, we present the time-history capturing the dynamics of optomechanical response of the EM when modulating the electromagnetic wave with the 1st and 2nd mechanical modes. This response reveals a periodic train of narrow pulses. Each pulse signifies the generation of an extensive array of frequency combs, showcasing the intricate interplay between mechanical and optical elements in the EM-cavity optomechanical setup. Notably, we observe that for the first mechanical mode case, the period of the pulse train is precisely determined to be 0.28 ms. This temporal characteristic aligns with the corresponding $FSR_1$ of 3.5 kHz, emphasizing the resonance and coherence achieved through optomechanical coupling. When the second mechanical mode is excited, the period $T_2$ is determined to be 0.11 ms, aligning perfectly with an $FSR_2 = 9$ kHz. Delving deeper into the intricacies of each period within the time-domain response, the shape of each period resembles that of a solitary wave, offering insights into the nonlinear and self-sustaining nature of the optomechanically induced FCs. This soliton-like behavior in the time domain underscores the stability and coherence achieved in the generation of the FCs, adding a layer of significance to the observed phenomena.

To validate our study, we introduced a second CW laser beam (726.65 kHz, 1mW). This laser beam interacted with the EM resonator, leading to the generation of optomechanical soliton FCs at lower frequency domain. This adjustment allowed for the possibility of interference between the generated solitons, facilitating the creation of a dual-frequency comb with a smaller FSR, thus achieving improved resolution. Here, the acoustic waveform is set to drive the EM's second mechanical mode. The lower the acoustic forcing intensity, the fewer the mechanical frequency combs appear, and a primary soliton emerges, centred at the first electromagnetic mode. However, as we increased the amplitude of the acoustic force to 110 dB, an intriguing interference pattern materialized. This interference occurred between a soliton centred at zero frequency and a secondary soliton centred at the first optical mode, as illustrated in Figure 6(c). Importantly, this interference led to the creation of a finer frequency spacing in the frequency domain, resulting in an FSR of 2 kHz. Significantly, this spacing can be effectively controlled, either increased or decreased, by manipulating either the mechanical mode of the EM or the optical mode of the laser source. The generation of these soliton FCs represents a notable breakthrough, distinguished by their remarkable simplicity and efficiency compared to existing implementations. This advancement opens the door to a multitude of applications, including spectroscopy and quantum computing, offering promising avenues for future research and technological development.

## Conclusion

This study introduces an innovative approach that leverages EM-optomechanical cavity technology to generate multi-soliton optomechanical FCs operating within the microwave frequency range, while establishing kHz-level spacing within a Kerr-comb medium. Here, a circular elastomer membrane is subjected to a different laser pump (power and frequency), and an acoustic wave generator precisely excites the membrane at its natural frequencies which modulates the electromagnetic wave of the laser source. This results in the creation of stable and resilient mode-locked multi-optomechanical soliton FCs, perfectly aligned with the acoustic wave frequency spacing. Furthermore, it enables the creation of dual frequency combs by merging the solitons, yielding a finer FSR that significantly enhances resolution. This innovative soliton frequency comb architecture holds immense fields such as metrology, quantum computing, and notably, sensing through spectroscopy by leveraging the concept of dual-frequency combs.

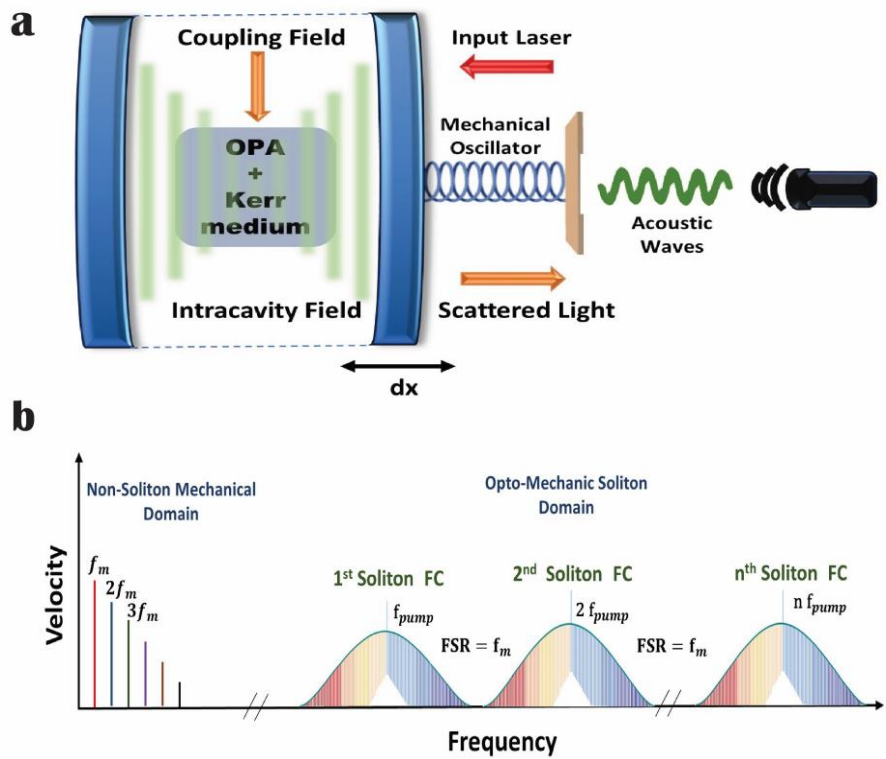

**Figure 1: Generation of Multi-Soliton Optomechanical FCs in an EM-Cavity Resonator. *a,*** *illustrates the mechanism employed to generate the multi-soliton FCs within an optomechanical EM-cavity resonator. The system features an elastomer membrane as the flexible mechanical component placed on an optical cavity. The EM resonator encounters a continuous-wave (CW) laser beam with constant optical power in two cases: Case 1 (5mW and $f_o$=3.505 MHz) and Case 2 (1mW and $f_o$=726.65 kHz). As the laser beam passes through the elastomer, it enters the cavity and is reflected from the gold-coated cavity floor, leading to the formation of counter-propagating photon pairs within the cavity. Excitation of the EM' motion (dx) is achieved by introducing a harmonic acoustic wave tuned to the EM's resonant frequencies, resulting in the generation of a nonlinear Kerr medium within the optical cavity. Amplification of the Kerr nonlinear coupling is achieved by increasing the acoustic force. This enhancement leads to, **b,** the emergence of multi-soliton optomechanical FCs centred at the laser pump frequency and its higher harmonics.*

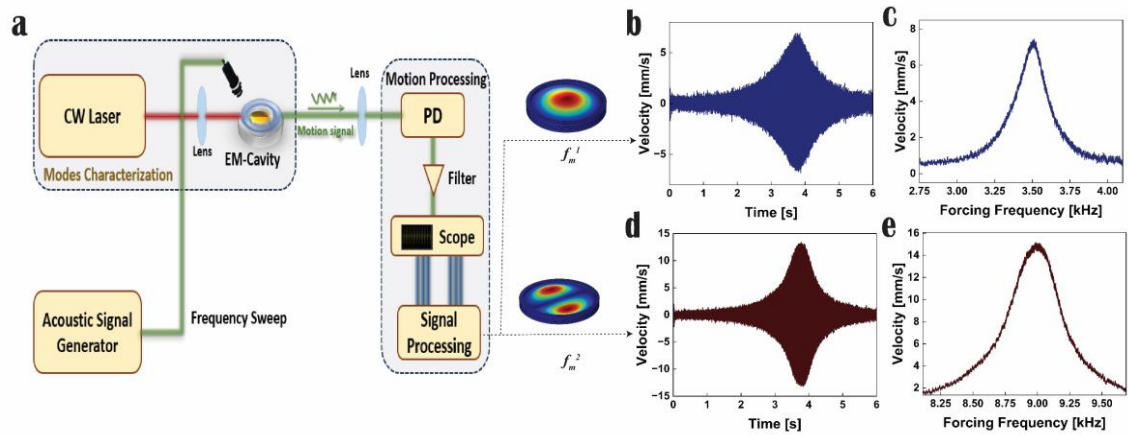

**Figure 2: Elastomer Membrane Characterization.** *a Schematic of the experimental setup used to measure the EM's optomechanical response. A CW Laser is employed to introduce an optical field inside the cavity and a Photodiode (PD) is used to measure the EM's optomechanical response.* ***b,*** *and* ***c,*** *Time-domain velocity responses of the EM's centre point as the acoustic frequency is swept within a range from 2.5 kHz to 4.5 kHz, and 8 kHz to 10 kHz, respectively, for 6 seconds* ***d,*** *and* ***e,*** *The EM's frequency-response curves obtained by calculating the RMS of the corresponding time-domain response.*

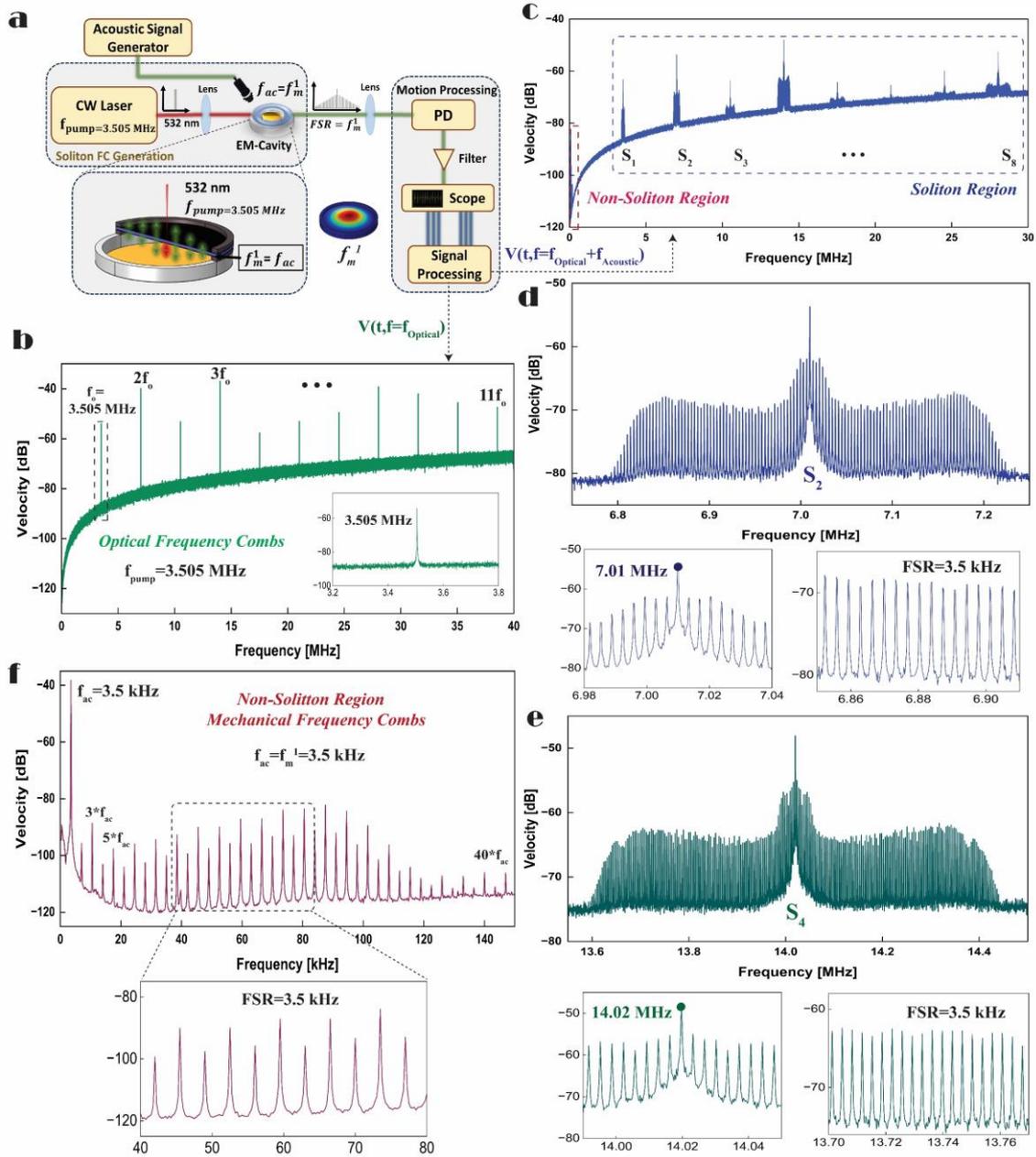

**Figure 3: Optomechanical Soliton FCs Generated via Exciting the EM's First Resonant Mode. a,** Experimental setup employed for soliton FCs generation. **b,** Frequency spectrum of the EM's central point velocity revealing eleven equally spaced dominant harmonics above the noise level. The primary peak at $f_o$=3.505 MHz corresponds to the laser pump frequency. **c,** Formation of multi soliton optomechanical FCs by directly exciting the second resonant mode of the elastomer membrane through an external acoustic signal ($f_{ac} = f_m^1 = 3.5\ kHz$). **d,** and **e,** Frequency spectra of the EM's optomechanical response corresponding to the second and fourth solitons, respectively. **f,** Mechanical FCs spanning from 0 to 150 kHz.

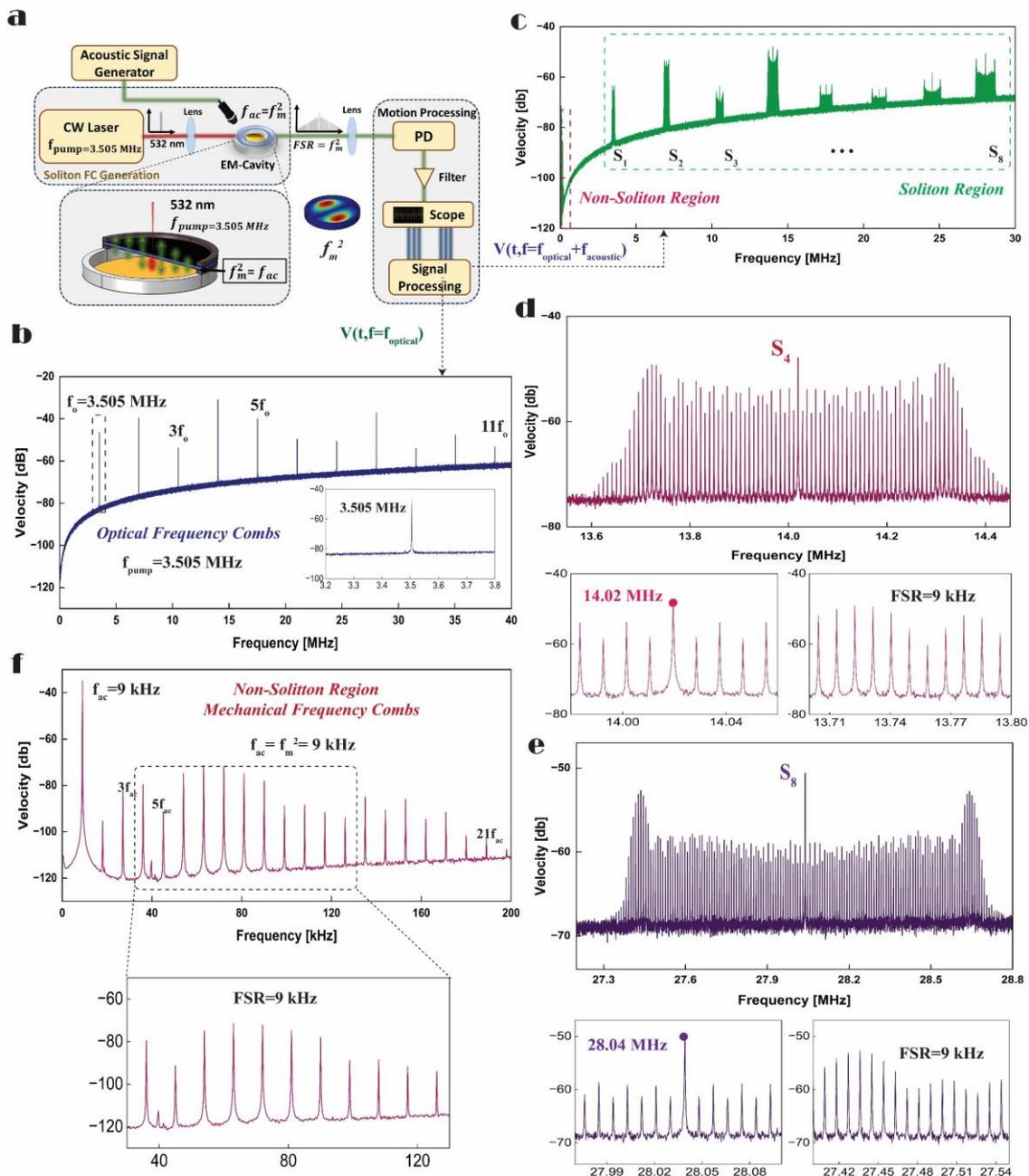

**Figure 4: Optomechanical Soliton Frequency Combs Generated via driving the EM's Second Resonant Mode.** *a,* *Experimental setup employed for soliton FCs generation.* *b,* *Frequency spectrum of the EM's central point velocity revealing eleven equally spaced dominant harmonics above the noise level.* *c,* *Formation of multi soliton optomechanical FCs by directly exciting the second resonant mode of the elastomer membrane using an external acoustic signal ($f_{ac} = f_m^2 = 9\ kHz$).* *d, and e,* *Frequency spectra of the EM's optomechanical response corresponding to the second and fourth solitons, respectively.* *f,* *Mechanical FCs spanning from 0 to 200 kHz.*

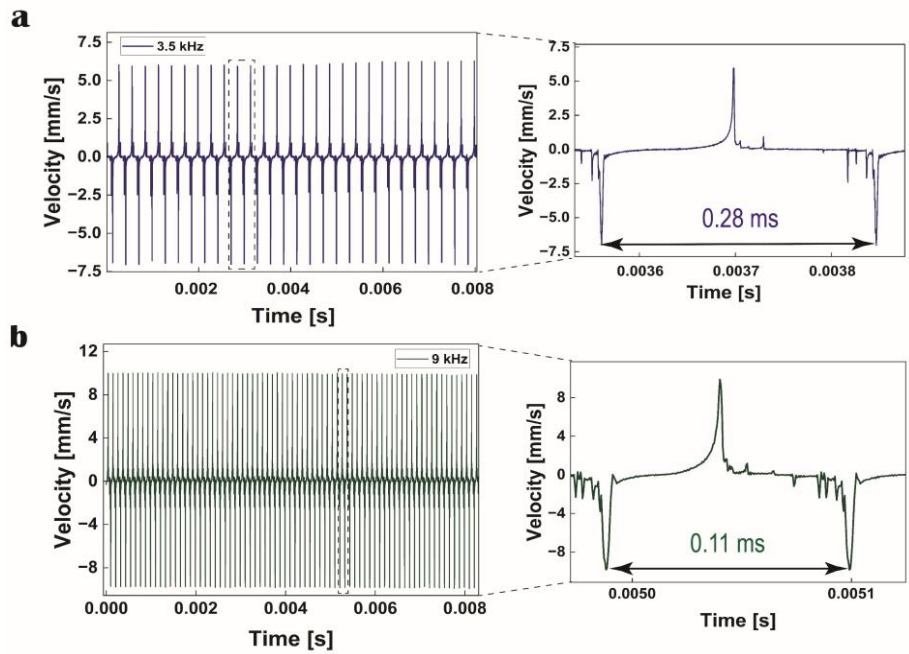

**Figure 5: Time-domain Analysis of the Optomechanical Response in the EM Resonator.** *a,* *Showing the 3.5 kHz case, with a magnified view highlighting one period ($T_1$ = 0.28 ms) and exhibiting a soliton shape.* *b,* *For the 9 kHz case, featuring a close-up perspective of one period ($T_2$ = 0.11 ms) and showcasing a soliton shape.*

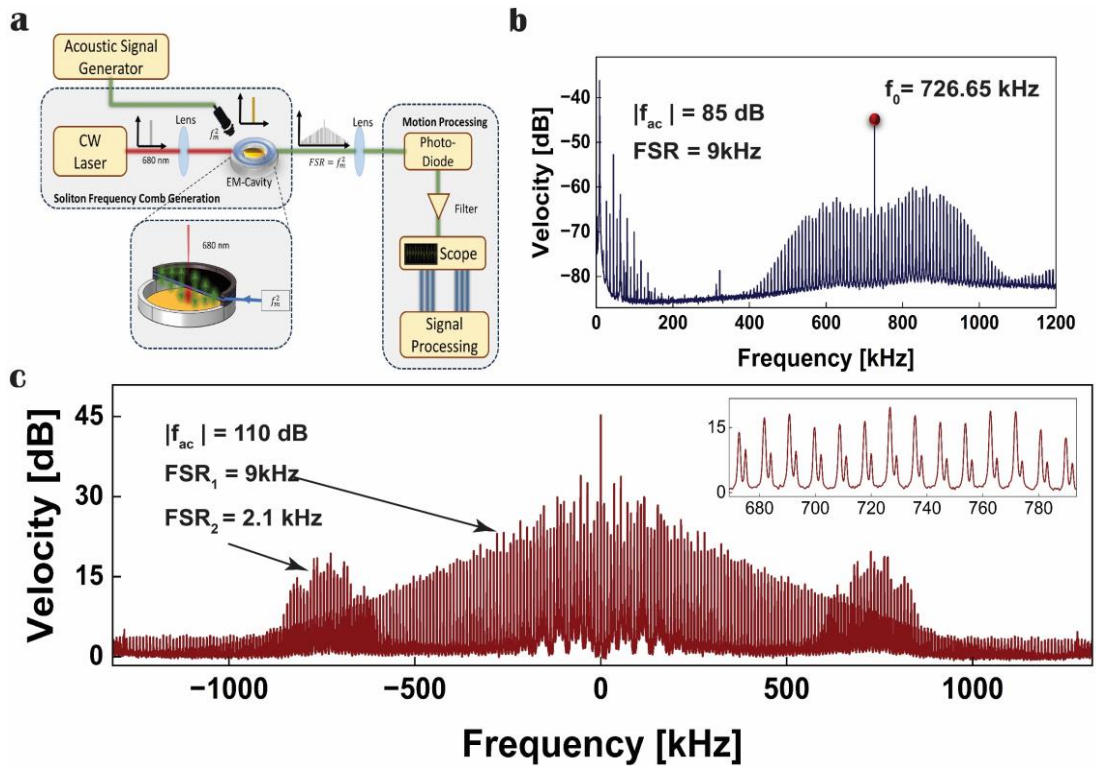

**Figure 6: Dual Frequency Combs Resulting from Interference between the Generated Solitons FCs.** *a, Experimental setup used for soliton FCs generation with a continuous-wave (CW) laser (680nm, 1mW, and $f_0$=726.65kHz). b, Mechanical FCs that are multiple of $f_{ac}$ ranging from 0 to 150 kHz and optomechanical FCs centred at $f_o$ with FSR=$f_{ac}$=9kHz. c, Interference between mechanical Soliton Frequency Combs (FCs) centred at zero and optomechanical FC centred at $f_o$ (for $|f_{ac}|$=110 dB), resulting in dual frequency combs with a finer FSR of 2.1 kHz.*

# Methods

**EM-Cavity Resonator Fabrication**

The elastomer membrane (EM)-cavity resonator was fabricated as outlined in Figure ED1. The optomechanical component, composed of VHB 4910 elastomer with a Young's modulus of 1.7 MPa and a 50 μm thickness, underwent the following steps: 1) Biaxial stretching generated pre-stress within the elastomer. 2) The stretched elastomer adhered to a rigid ring. 3) A partially covering layer of 846-80G Carbon Grease on the membrane's underside traps photons, enabling laser transmission, while a 1 μm gold coating on the cavity floor ensured strong laser beam reflection. 4) A separating paper introduced a customized gap between the elastomer and Gold-PMMA substrate. It is worth noting that this is a customized device we fabricated manually in our Lab.

**Optomechanical Soliton FC Generation: Frequency Detuning**

Our experimental findings establish a direct link between the intensity of soliton FCs and the frequency detuning, $\delta = f_{ac} \pm f_m^i$ $(i = 1, 2, \ldots)$, between the acoustic wave and mechanical resonant modes of the EM resonator. Efficient soliton formation occurs when the acoustic wave frequency aligns precisely with the mechanical natural frequency of the EM resonator. As the detuning between the acoustic wave and mechanical mode frequencies increases, the strength of soliton combs diminishes. Figure ED3 showcases velocity frequency spectra of the EM resonator under two acoustic frequency detuning scenarios: 1) acoustic wave frequency slightly below the EM's first mode natural frequency ($\delta = f_{ac} - f_m^1 < 0$ ), illustrated in Figures ED3 (a, b, c, and d), 2) acoustic wave driving the EM's second mode from above ($\delta = f_{ac} - f_m^2 > 0$), shown in Figures ED3 (d, e, and f). We conclude that higher mechanical resonant modes offer a greater margin of stability for the generation of the proposed Kerr-cavity solitons.

**Acknowledgements** We acknowledge the funding from the European Union's Horizon 2020 Marie Sktodowska-Curie Actions -Innovative European Training Networks under grant agreement No 956401.

**Author contributions**

Conceptualization: SR, HM

Methodology: SR, HM, AA, ZI

Mathematical modelling: SR, AA

Numerical simulation: SR

Designing experiments: HM, SR

Device fabrication: HM, AA, SR

Investigation: SR, HM, AA, ZI

Visualization: HM, SR

Supervision: FJM, LSM, JA

Writing—original draft: SR, HM, ZI, AA

Writing—review & editing: HM, SR, AA, ZI, FJM, LSM, JA

# Extended Data

## ED 1: Fabrication of the EM-Cavity Resonator

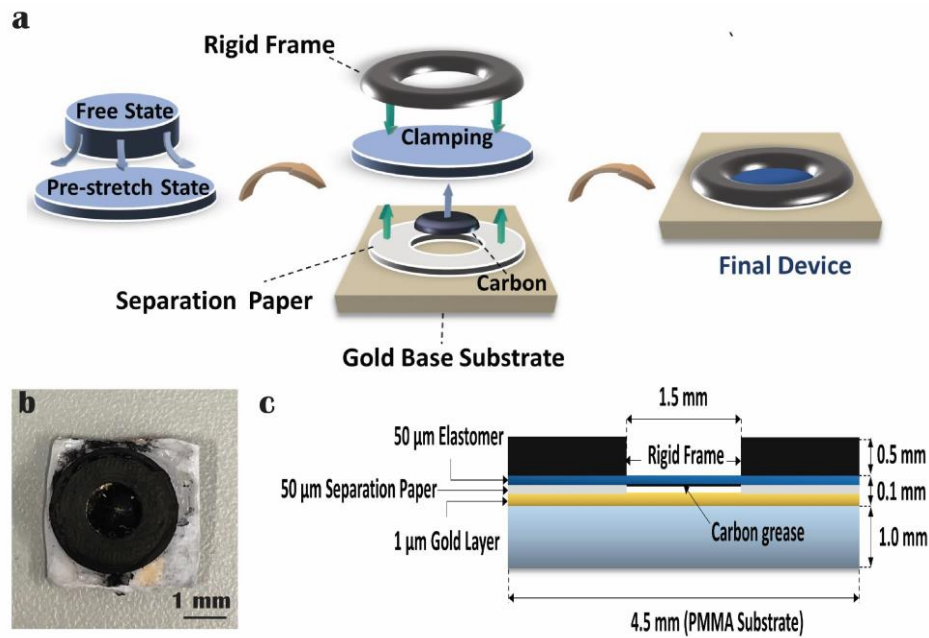

**Extended Data Figure 1:** *Fabrication of the EM-cavity. a*, Shows a 5cm×5cm elastomer piece that is cut from an elastomer roll of type VHB 4910. The elastomer's thickness is 1mm in its unstretched configuration. The elastomer is stretched along two perpendicular directions to induce a pre-stretch mechanical load within the material, tuning the membrane's natural frequencies and transmissibility. The stretched elastomer is then glued to a rigid ring forming fixed boundary conditions. Afterwards, the elastomer's bottom surface is coated with a thin layer of carbon grease of type 846-80G (not totally covering the surface). Moreover, the cavity floor is coated with a thin layer of gold (1 µm), providing strong reflection for the incident laser beam. A perforated separating paper is inserted between the elastomer membrane and the PMMA substrate to create a gap between them and form the cavity system. *b,* The fabricated EM-cavity device. *c,* The device's cross-sectional view showing the dimensions of the layered structure and its hierarchy.

## ED 2: Mechanical Time History of the EM Resonator.

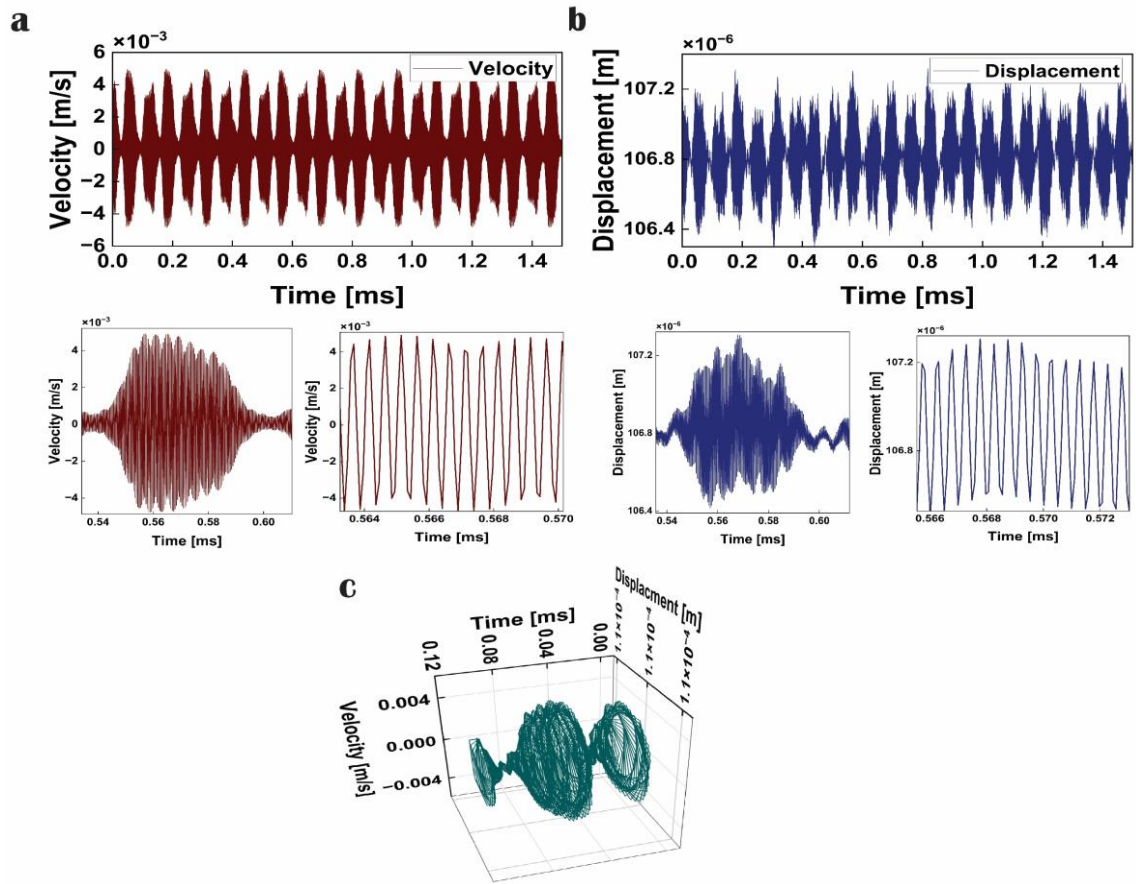

**Extended Data Figure 2:** *Measured EM's time-history under acoustic waveform.*
*The figure illustrates the EM's response measured within a time window of 1.45 ms that contains only the mechanical frequency combs shown in Figure 2f, where an acoustic waveform with amplitude of 105 dB is set to directly drive the EM's first out-of-plane mode, $f_{ac} = f_m^1 = 3.5$ kHz. **a,** The EM centre point displacement together with the zoomed areas showing the response envelope in more detail. **b,** The EM's trajectory shown in a 3D phase-space diagram.*

## ED 3: Frequency Detuning

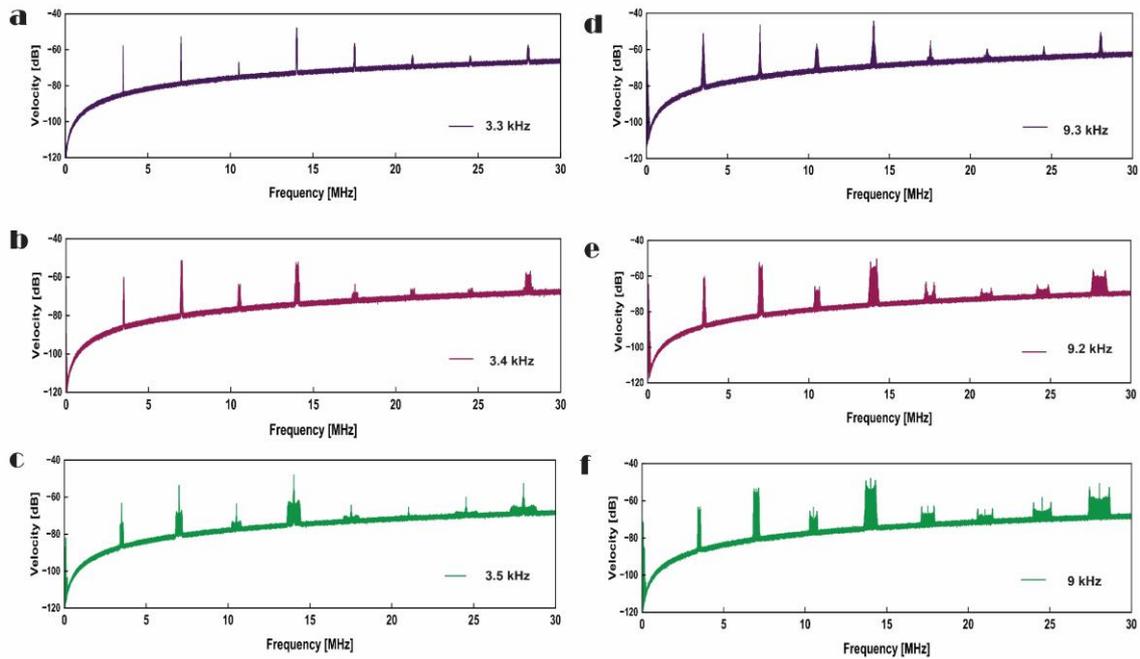

**Extended Data Figure 3: The influence of *frequency detuning on the solitons' formation. a,*** *The frequency of the acoustic wave (OPA) is detuned "below" the EM's first natural frequency by 0.2 kHz, $\delta = f_{ac} - f_m^1 = -0.2\ kHz$, which does not create soliton frequency combs.* ***b,*** *Reducing the detuning parameter down to $\delta = f_{ac} - f_m^1 = -0.1\ kHz$ triggers the formation of narrowband solitons centered at integer harmonics of the laser pump frequency.* ***c,*** *Depicts the generation of mode-locked multi soliton frequency combs for the case of exact syntonization between the acoustic signal and the EM's first mechanical mode frequencies, $\delta = f_{ac} - f_m^1 = 0$.* ***d,*** *The acoustic wave frequency is detuned "above" the membrane's second natural frequency by 0.3 kHz, $\delta = f_{ac} - f_m^2 = 0.3\ kHz$. As seen, detuning from "above" can still provide a ghost of soliton in the optomechanical response.* ***e,*** *Reducing the detuning parameter down to $\delta = f_{ac} - f_m^2 = 0.2\ kHz$, the evolution of solitons formed around the integer harmonics of the pump frequency is strengthened.* ***f,*** *Zero detuning parameter, $\delta = f_{ac} - f_m^2 = 0$, leads to the highest power for the mode-locked solitons frequency combs.*

# Soliton Frequency Combs in Elastomer Membrane-Cavity Optomechanics: Supplementary Information

Sasan Rahmanian[1], Hamza Mouharrar[1], Amin Alibakhshi[2], Zeeshan Iqbal[1,3], Luis Saucedo-Mora[2,4,5], Francisco Javier Montáns[2,6], and Jan Awrejcewicz [7†]

[1]Department of Systems Design Engineering, University of Waterloo, Waterloo, ON N2L 3G1, Canada

[2]Escuela Técnica Superior de Ingeniería Aeronáutica y del Espacio, Universidad Politécnica de Madrid, Pza.Cardenal Cisneros, 28040 Madrid, Spain

[3]Department of Physics, Government College University, Katchery Road, Lahore 54000, Pakistan

[4]Department of Materials, University of Oxford, Parks Road, Oxford, OX1 3PJ, UK

[5]Department of Nuclear Science and Engineering, Massachusetts Institute of Technology, MA 02139, USA

[6]Department of Mechanical and Aerospace Engineering, Herbert Wertheim College of Engineering, University of Florida, Gainesville, FL, 32611, USA

[7]Department of Automation, Biomechanics and Mechatronics, Lodz University of Technology, 1/15 Stefanowskiego St., 90-924, Lodz, Poland

*Sasan Rahmanian and Hamza Mouharrar have equally contributed to this work.

## S1: Mathematical Model and Numerical Simulations

The circular elastomer membrane undergoes deformation when it is externally excited using optical and acoustic excitations. The derivation of the equations of motion describing the elastomer's dynamic behavior is established based on the generalized Hamilton's principle. Because the membrane is thin, Kirchhoff–Love plate theory is employed to form the displacement field governing the continuum's deformation[1]. Based on this theory, each material particle of the elastomer membrane displaces according to the following field in a cylindrical Lagrangian coordinates system, as given below,

$$U_r(r,\theta,z;t) = u(r,\theta;t) - z\,\frac{\partial w(r,\theta;t)}{\partial r}$$

$$U_\theta(r,\theta,z;t) = v(r,\theta;t) - \frac{z}{r}\,\frac{\partial w(r,\theta;t)}{\partial \theta}$$

$$\text{(ES1)}$$

$$U_z(r,\theta,z;t) = w(r,\theta;t)$$

where $U_r$, $U_\theta$, and $U_z$ stand for the radial, circumferential, and transverse displacement components, respectively. The nonlinear strain-displacement relationships are used to account for geometrical nonlinearities resulting from large deformations. The non-zero components of the Cauchy green strain tensor, including von Karman nonlinearities are given in equation (ES2).

$$\varepsilon_{rr} = \frac{\partial u}{\partial r} - z\,\frac{\partial^2 w}{\partial r^2} + \frac{1}{2}\left(\frac{\partial w}{\partial r}\right)^2$$

$$\varepsilon_{\theta\theta} = \frac{u}{r} + \frac{1}{r}\frac{\partial v}{\partial \theta} + \frac{1}{2r^2}\left(\frac{\partial w}{\partial \theta}\right)^2 - \frac{z}{r}\left(\frac{\partial w}{\partial r} - \frac{1}{r}\frac{\partial^2 w}{\partial \theta^2}\right)$$

$$\varepsilon_{r\theta} = \frac{1}{2}\left(\left(\frac{1}{r}\frac{\partial u}{\partial \theta} + \frac{\partial v}{\partial r} - \frac{v}{r} + \frac{1}{r}\frac{\partial w}{\partial r}\frac{\partial w}{\partial \theta}\right) - \frac{2z}{r}\left(\frac{\partial^2 w}{\partial r \partial \theta} - \frac{1}{r}\frac{\partial w}{\partial \theta}\right)\right)$$

$$\text{(ES2)}$$

$$\varepsilon_{zz} \neq 0, \varepsilon_{rz} = \varepsilon_{\theta z} = 0$$

To satisfy the incompressibility condition existing in structures that are made from hyperelastic materials, the strain component along the thickness direction, $\varepsilon_{zz}$, is non-zero and obtained as follows[2,3],

$$\varepsilon_{zz} = \frac{1}{2\left((2\varepsilon_{rr}+1)(2\varepsilon_{\theta\theta}+1) - \varepsilon_{r\theta}^2\right)} - \frac{1}{2} \qquad \text{(ES3)}$$

To simplify the problem, we truncate equation (ES3) around the zero equilibrium and retain terms up to quadratic nonlinearity, resulting in the following expression:

$$\varepsilon_{zz} \approx -(\varepsilon_{rr} + \varepsilon_{\theta\theta}) + 2\left(\varepsilon_{rr}^2 + \varepsilon_{\theta\theta}^2 + \varepsilon_{rr}\varepsilon_{\theta\theta}\right) + \frac{\varepsilon_{r\theta}^2}{2} \qquad \text{(ES4)}$$

We obtain the strain potential energy stored in the elastomer medium using a hyperelastic constitutive equation, the neo-Hookean model. Before obtaining the potential energy, we first need to introduce the right Cauchy-Green deformation tensor as

$$\boldsymbol{C} = \begin{pmatrix} 2\varepsilon_{rr} + 1 & \varepsilon_{r\theta} & 0 \\ \varepsilon_{r\theta} & 2\varepsilon_{\theta\theta} + 1 & 0 \\ 0 & 0 & 2\varepsilon_{zz} + 1 \end{pmatrix} \qquad (ES5)$$

To formulate the neo-Hookean-based strain potential energy, we require the first invariant of the right Cauchy-Green deformation tensor, which is obtained from equation (ES6) as

$$I_1 = 2(\varepsilon_{rr} + \varepsilon_{\theta\theta} + \varepsilon_{zz}) \qquad (ES6)$$

The neo-Hookean strain potential energy function is given as

$$\psi = \frac{\mu}{2}(I_1 - 3) \qquad (ES7)$$

in which $\mu$ is a material constant that is obtained from experiments. Eventually, the potential energy stored in the elastomer membrane due to its deformation is expressed as

$$E_S = \int_0^a \int_0^{2\pi} \int_{-h/2}^{h/2} \psi \, r dz d\theta dr \qquad (ES8)$$

The kinetic energy of the structure counting for linear translational inertias can be written in the form of

$$E_K = \frac{1}{2} \int_0^a \int_0^{2\pi} \int_{-h/2}^{h/2} \rho \left( \left(\frac{\partial u}{\partial t}\right)^2 + \left(\frac{\partial v}{\partial t}\right)^2 + \left(\frac{\partial w}{\partial t}\right)^2 \right) dz \, r dr \, d\theta \qquad (ES9)$$

where $\rho$ refers to the mass density of the hyperelastic plate. The variation of the work done by the externally applied force (harmonic acoustic wave) is given as

$$\delta W_e = \int_0^a \int_0^{2\pi} f_{ac}(r,\theta) \, cos(\omega t) \, \delta w \, r d\theta dr \qquad (ES10)$$

where, $f_{ac}(r, \theta)$ is the amplitude of the acoustic pressure observed at the EM's top surface, and $\omega$ is the acoustic wave's angular frequency. The variation of the work done by the nonconservative viscous damping force is formulated as

$$\delta W_d = -c_d \int_0^a \int_0^{2\pi} \left( \frac{\partial u}{\partial t} \delta u + \frac{\partial v}{\partial t} \delta v + \frac{\partial w}{\partial t} \delta w \right) r d\theta dr \qquad (ES11)$$

in which $c_d$ is the damping coefficient. We exploited the generalized Hamilton's principle, i.e., $\delta \int_{t_1}^{t_2}(E_K - E_S)dt + \delta \int_{t_1}^{t_2}(W_e + W_d)\, dt = 0$ to derive the system's nonlinear motion's equations. In this context, the membrane is assumed to be operating in the vicinity of the first few out-of-plane/transverse modes. It is important to note that the membrane is almost stationary along the circumferential direction within this frequency range, meaning, $v(r, \theta; t) = 0$. Furthermore, because the structure is symmetric, the natural frequencies of the membrane's non-axisymmetric radial modes are significantly higher than the operating frequency, rendering them negligible in contributing to the membrane's radial motion. Defining the following dimensionless parameters

$$\tilde{u} = \frac{u}{h} \; , \tilde{w} = \frac{w}{h} \; , \qquad \bar{r} = \frac{r}{a} \; , \; \tilde{t} = \frac{t}{t^*} \; \text{ where } \; t^* = \sqrt{\frac{\rho h a^4}{D_E}} \qquad (ES12)$$

a set of the nonlinear partial differential equations (PDEs) describing the circular EM's in-plane radial and out-of-plane motions are expressed in the following non-dimensional form (the tildes are dropped for the sake of simplicity),

$$\delta u: \quad \frac{\partial^2 u}{\partial t^2} + c_u \frac{\partial u}{\partial t} + 6\beta_1 \frac{u}{r^2} - 4\beta_1 \frac{\partial^2 u}{\partial r^2} + \beta_2 \frac{1}{r}\left(\frac{\partial w}{\partial r}\right)^2 + 4\beta_2 \frac{1}{r^3}\left(\frac{\partial w}{\partial \theta}\right)^2$$
$$- 4\beta_2 \left(\frac{\partial w}{\partial r}\right)\left(\frac{\partial^2 w}{\partial r^2}\right) - \frac{\beta_2}{4}\frac{1}{r^2}\left(\frac{\partial^2 w}{\partial \theta^2}\right)\left(\frac{\partial w}{\partial r}\right) - \frac{9\beta_2}{4}\frac{1}{r^2}\frac{\partial w}{\partial \theta}\frac{\partial^2 w}{\partial r \partial \theta} = 0 \qquad (ES13)$$

$\delta w$:  
$$\frac{\partial^2 w}{\partial t^2} + c_w \frac{\partial w}{\partial t} - 16\sigma_1 \frac{1}{r^4} \frac{\partial^2 w}{\partial \theta^2} + 4\sigma_1 \frac{1}{r^4} \frac{\partial^4 w}{\partial \theta^4} + 12\sigma_1 \frac{1}{r^3} \frac{\partial w}{\partial r}$$
$$- 6\sigma_1 \frac{1}{r^2} \frac{\partial^2 w}{\partial r^2} + 4\sigma_1 \frac{1}{r^4} \frac{\partial^4 w}{\partial r^4} - \frac{5}{4}\sigma_2 \frac{1}{r^2} \left(\frac{\partial^2 w}{\partial \theta^2}\right)\left(\frac{\partial w}{\partial r}\right)^2$$
$$- 6\sigma_2 \frac{1}{r^4} \left(\frac{\partial w}{\partial \theta}\right)^2 \left(\frac{\partial^2 w}{\partial \theta^2}\right) - 4\sigma_3 \frac{1}{r} u \frac{\partial^2 w}{\partial \theta^2} - 2\sigma_3 \frac{1}{r^2} \left(\frac{\partial^2 w}{\partial \theta^2}\right)\left(\frac{\partial u}{\partial r}\right)$$
$$+ \frac{5}{2}\sigma_2 \frac{1}{r^3} \left(\frac{\partial w}{\partial \theta}\right)^2 \left(\frac{\partial w}{\partial r}\right) + 2\sigma_3 \frac{1}{r^2} u \left(\frac{\partial w}{\partial r}\right)$$
$$- 5\sigma_2 \frac{1}{r^2} \left(\frac{\partial w}{\partial \theta}\right)\left(\frac{\partial w}{\partial r}\right)\left(\frac{\partial^2 w}{\partial r \partial \theta}\right) - 4\sigma_3 \left(\frac{\partial w}{\partial r}\right)\left(\frac{\partial^2 u}{\partial r^2}\right) \qquad (ES14)$$
$$- \frac{5}{4}\sigma_2 \frac{1}{r^2} \left(\frac{\partial w}{\partial \theta}\right)^2 \left(\frac{\partial^2 w}{\partial r^2}\right) - 6\sigma_2 \left(\frac{\partial w}{\partial r}\right)^2 \left(\frac{\partial^2 w}{\partial r^2}\right)$$
$$- 2\sigma_3 \frac{1}{r} u \left(\frac{\partial^2 w}{\partial r^2}\right) - 4\sigma_3 \left(\frac{\partial u}{\partial r}\right)\left(\frac{\partial^2 w}{\partial r^2}\right) = F_{ac} \cos(\Omega t)$$

where, the equations' dimensionless coefficients are given as

$$\beta_1 = \frac{\mu h a^2}{D_E} \ , \ \ \beta_2 = \frac{\mu a h^2}{D_E} \ , \ \ c_u = \frac{\tilde{c}_u}{d_1} \sqrt{\frac{h a^4}{\rho D_E}}$$
$$\sigma_1 = \frac{\mu h d_2}{D_E d_1}, \ \ \sigma_2 = \frac{\mu h^3}{D_E} \ , \qquad \sigma_3 = \frac{\mu a h^2}{D_E} \ , c_w = \frac{\tilde{c}_w}{d_1} \sqrt{\frac{h a^4}{\rho D_E}} \ , \qquad (ES15)$$
$$F_{ac} = \frac{a^4 f_{ac}(r, \theta)}{D_E d_1} \ , \ \ \Omega = \omega t^*$$

Subject to the following boundary conditions.

$$u(r, \theta; t) = w(r, \theta; t) = 0 \ @ \ r = 1$$
$$\frac{\partial w}{\partial r}(r, \theta; t) = 0 \ @ r = 1 \qquad (ES16)$$

In equations (ES13), (ES14), and (ES15), $d_1$ and $d_2$ are defined as

$$d_1 = \int_{-h/2}^{h/2} dz = h \ \ , \qquad d_2 = \int_{-h/2}^{h/2} z^2 \, dz = \frac{h^3}{12} \qquad (ES17)$$

The two coupled PDEs introduced in equations (ES13) and (ES14) are nonlinear and almost impossible to present a closed-form solution to them. It is common to discretize the PDEs into a set of nonlinearly coupled ordinary differential equations (ODEs) via implementing Galerkin's method. To this, the membrane's radial and transverse displacements are expanded as a linear combination of the comparison functions, $\phi_i$ and $\psi_{nm}$, satisfying all the system's boundary conditions, as

$$u(r,\theta;t) = \sum_{i=1}^{N_u} p_i(t)\phi_i(r) \quad , \quad w(r,\theta;t) = \sum_{n=1}^{N}\sum_{m=0}^{M} q_{nm}(t)\psi_{nm}(r,\theta) \qquad (ES18)$$

where, $p_i$ and $q_{nm}$ are, respectively, the generalized coordinates of the in-plane radial and out-of-plane motions. $N_u$ is the number of modes considered for the in-plane radial motion. $N$ and $M$ are the maximum number assumed for the first and second index of the transverse mode shapes, so $N \times (M+1)$ is the number of modes considered to expand the out-of-plane displacement. Here, $N_u = 1$, $N = 2$, and $M = 1$. $\phi_i$s are the eigenfunctions of the linear undamped system corresponding to equation (ES12), satisfying:

$$4\beta_1 r^2 \phi_i^{''} + \left(\lambda_i^2 r^2 - 6\beta_1\right)\phi_i = 0 \qquad (ES19)$$

subject to $\phi(1) = 0$ and $\phi(0) =$ finite.

It is noteworthy that equation (ES19) is convertible to a Bessel equation and the solution to the obtained Bessel-type equation (ES19) is written in terms of the Bessel function of the first kind as $\phi_i(r) = A_i\sqrt{r}J_{\frac{\sqrt{7}}{2}}\left(\sqrt{\frac{1}{4\beta_1}}\lambda_i r\right)$, where, $\lambda_i$s are the roots of the following characteristic equation: $J_{\frac{\sqrt{7}}{2}}\left(\sqrt{\frac{1}{4\beta_1}}\lambda_i\right) = 0$, $(i = 1,2, \dots)$. $\phi_i$s is mass-normalized such that $\int_0^1 \phi_i^2 dr = \frac{1}{2\pi}$. Moreover, $\psi_{nm}$s are the eigenfunctions of a linear elastic thin circular plate under clamped boundary conditions, given as[4].

$$\psi_{nm}(r,\theta) = \left(J_m(\gamma_{nm}r) - \frac{J_m(\gamma_{nm})}{I_m(\gamma_{nm})}I_m(\gamma_{nm}r)\right)(\cos m\theta + \sin m\theta) \qquad (ES20)$$

here, $\beta_n$s are the roots of the having the following characteristics equation[5].

$$I_m(\gamma_{nm})J_{m-1}(\gamma_{nm}) - J_m(\gamma_{nm})I_{m-1}(\gamma_{nm}) = 0 \;\; ; \;\; n = 1,2, \dots, m = 0,1,2, \dots \qquad (ES21)$$

Introducing equations. (ES18) into equations. (ES13) and (ES14), and employing the Galerkin's technique, a system of the nonlinear ODEs governing the EM's dynamics, which is the expanded form of equations (2) and (3), is obtained in the form of:

$$\ddot{p}_n + c_u \dot{p}_n + \lambda_n^2 p_n + \sum_{i=1}^{N} \sum_{j=0}^{M} \sum_{k=1}^{N} \sum_{l=0}^{M} K_{nijkl}^{u(q)} q_{ij} q_{kl} = 0 \tag{ES22}$$

$$\ddot{q}_{nm} + c_w \dot{q}_{nm} + \sum_{i=1}^{N} \sum_{j=0}^{M} K_{nmij}^{w(l)} q_{ij} + \sum_{i=1}^{N_u} \sum_{k=1}^{N} \sum_{l=0}^{M} K_{nmikl}^{w(q)} p_i q_{kl}$$
$$+ \sum_{i=1}^{N} \sum_{j=0}^{M} \sum_{k=1}^{N} \sum_{l=0}^{M} \sum_{s=1}^{N} \sum_{t=0}^{M} K_{nmijklst}^{w(c)} q_{ij} q_{kl} q_{st} \tag{ES23}$$
$$= \cos(\Omega t) f_{nm} + \frac{\Gamma}{\left(1 + Rw(0,0;t)\right)^2} |\alpha(t)|^2$$

where the components of the in-plane's quadratic stiffness, $K_{nijkl}^{u(q)}$, and the out-of-plane linear, quadratic, and cubic stiffnesses, $K_{nmij}^{w(l)}$, $K_{nijkl}^{u(q)}$, and $K_{nmijklst}^{w(c)}$ are defined as follows.

$$K_{nijkl}^{u(q)} = \beta_2 \int_0^1 \int_0^{2\pi} \frac{1}{r} \phi_n \psi_{ij,r} \psi_{kl,r} dr \, d\theta + 4\beta_2 \int_0^1 \int_0^{2\pi} \frac{1}{r^3} \phi_n \psi_{ij,\theta} \psi_{kl,\theta} dr \, d\theta$$
$$- 4\beta_2 \int_0^1 \int_0^{2\pi} \phi_n \psi_{ij,r} \psi_{kl,rr} dr \, d\theta$$
$$- \frac{\beta_2}{4} \int_0^1 \int_0^{2\pi} \frac{1}{r^2} \phi_n \psi_{ij,\theta\theta} \psi_{kl,r} dr \, d\theta$$
$$- \frac{9\beta_2}{4} \int_0^1 \int_0^{2\pi} \frac{1}{r^2} \phi_n \psi_{ij,\theta} \psi_{kl,r\theta} dr \, d\theta$$

$$K_{nmij}^{w(l)} = -16\sigma_1 \int_0^1 \int_0^{2\pi} \frac{1}{r^4} \psi_{nm} \psi_{ij,\theta\theta} \, r \, dr \, d\theta \tag{ES24}$$
$$+ 4\sigma_1 \int_0^1 \int_0^{2\pi} \frac{1}{r^4} \psi_{nm} \psi_{ij,\theta\theta\theta\theta} \, r \, dr \, d\theta$$
$$+ 12\sigma_1 \int_0^1 \int_0^{2\pi} \frac{1}{r^3} \psi_{nm} \psi_{ij,r} \, r \, dr \, d\theta$$
$$- 6\sigma_1 \int_0^1 \int_0^{2\pi} \frac{1}{r^2} \psi_{nm} \psi_{ij,rr} \, r \, dr \, d\theta$$
$$+ 4\sigma_1 \int_0^1 \int_0^{2\pi} \psi_{nm} \psi_{ij,rrrr} \, r \, dr \, d\theta$$

$$K_{nmikl}^{w(q)} = -4\sigma_3 \int_0^1 \int_0^{2\pi} \frac{1}{r^3} \psi_{nm} \phi_i \psi_{kl,\theta\theta} \; r \, dr \, d\theta$$
$$-2\sigma_3 \int_0^1 \int_0^{2\pi} \frac{1}{r^2} \psi_{nm} \phi_i' \psi_{kl,\theta\theta} \; r \, dr \, d\theta$$
$$+2\sigma_3 \int_0^1 \int_0^{2\pi} \frac{1}{r^2} \psi_{nm} \phi_i \psi_{kl,r} \; r \, dr \, d\theta$$
$$-4\sigma_3 \int_0^1 \int_0^{2\pi} \psi_{nm} \phi_{i,rr} \psi_{kl,r} \; r \, dr \, d\theta$$
$$-2\sigma_3 \int_0^1 \int_0^{2\pi} \frac{1}{r} \psi_{nm} \phi_i \psi_{kl,rr} \; r \, dr \, d\theta$$
$$-4\sigma_3 \int_0^1 \int_0^{2\pi} \psi_{nm} \phi_{i,r} \psi_{kl,rr} \; r \, dr \, d\theta$$

$$K_{nmijklst}^{w(c)} = -\frac{5}{4}\sigma_2 \int_0^1 \int_0^{2\pi} \frac{1}{r^2} \psi_{nm} \psi_{ij,\theta\theta} \psi_{kl,r} \psi_{st,r} \; r \, dr \, d\theta$$
$$-6\sigma_2 \int_0^1 \int_0^{2\pi} \frac{1}{r^4} \psi_{nm} \psi_{ij,\theta} \psi_{kl,\theta} \psi_{st,\theta\theta} \; r \, dr \, d\theta$$
$$+\frac{5}{2}\sigma_2 \int_0^1 \int_0^{2\pi} \frac{1}{r^3} \psi_{nm} \psi_{ij,\theta} \psi_{kl,\theta} \psi_{st,r} \; r \, dr \, d\theta$$
$$-5\sigma_2 \int_0^1 \int_0^{2\pi} \frac{1}{r^2} \psi_{nm} \psi_{ij,\theta} \psi_{kl,r} \psi_{st,r\theta} \; r \, dr \, d\theta$$
$$-\frac{5}{4}\sigma_2 \int_0^1 \int_0^{2\pi} \frac{1}{r^2} \psi_{nm} \psi_{ij,\theta} \psi_{kl,\theta} \psi_{st,rr} \; r \, dr \, d\theta$$
$$-6\sigma_2 \int_0^1 \int_0^{2\pi} \psi_{nm} \psi_{ij,r} \psi_{kl,r} \psi_{st,rr} \; r \, dr \, d\theta$$

$$f_{nm} = \int_0^1 \int_0^{2\pi} \psi_{nm} F_{ac} \; r \, dr \, d\theta$$

The optomechanical cavity illustrated in Figure 1 consists of two mirrors: a fixed mirror made of gold, which is coated on the substrate, and a suspended mirror formed by the bottom surface of the elastomer membrane. The suspended EM resonator can move in response to an optical force. This setup creates an optical cavity where light can bounce back and forth between the two mirrors, leading to the interaction between light and the mechanical motion of the elastomer membrane. For a relatively large mechanical out-of-plane displacement $w(r, \theta; t)$, the optical resonance frequency can be written as $\omega_{opt} = \frac{\pi c}{L_0 + w}$, where $c$ is the speed of light, and $L_0$ is the initial length of the optical cavity. The total optical energy $U_{opt}$ in the cavity can be written in terms of the modal amplitude $\alpha$ of the optical mode, which is normalized such that $|\alpha|^2$ represents the number of intracavity photons per unit area of the membrane. The expression for the optical energy density per unit area of the membrane is $U_{opt} = \hbar\,\omega_{opt}|\alpha|^2$, where $\hbar$ is the reduced Planck constant. For finite out-of-plane displacement of the elastomer membrane, the optical force applied to per unit area of the structure is: $F_{opt}(r,\theta;t) = -\partial U_{opt}/\partial w$ introducing a displacement-dependent optical force that is applied to the hyperelastic membrane along its transverse motion. This force is represented by the second term appearing in the right-hand side of equation (S23). In this equation, $\Gamma = \frac{a^4 \hbar \pi c}{D_E d_1 L_0^2}$ and $R = \frac{h}{L_0}$. It is assumed that the input photon flux is lumped at the membrane centre defined by the following non-dimensional dynamics:

$$\frac{d\alpha}{dt} = \big(i(\omega_L - \omega(w)) - \chi_1\big)\alpha + \chi_2 \qquad (ES25)$$

here, $\omega_L$ is the laser pump frequency, and $\omega(w) = t^*\omega_{opt}$ denotes the non-dimensional optical resonance frequency; $\chi_1 = \frac{\kappa t^*}{2}$ and $\chi_2 = \sqrt{\kappa_e}\,s_{in}t^*$, where $\kappa = \kappa_e + \kappa_l$ represents the sum of external and internal optical losses, $s_{in}$ refers to the input photon flux. The established equations of motion can be re-written in the following matrix form.

$$\ddot{\mathbf{p}} + c_u\,\dot{\mathbf{p}} + \lambda^2\mathbf{p} + (\mathbb{K}^u : \mathbf{Q}) : \mathbf{Q} = \mathbf{0} \qquad (ES26)$$

$$\ddot{\mathbf{Q}} + c_w\dot{\mathbf{Q}} + \mathbb{M} : \mathbf{Q} + (\mathbb{N}.\mathbf{p}) : \mathbf{Q} + \big((\mathbb{L} : \mathbf{Q}) : \mathbf{Q}\big) : \mathbf{Q}$$
$$= \cos(\omega t)\mathbf{f} + \frac{\Gamma}{\big(1 + Rw(0,0;t)\big)^2}|\alpha(t)|^2 \qquad (ES27)$$

Where $\mathbf{f}$ is the external force vector; $\mathbf{p}$ is a vector (tensor of rank 1); $\mathbf{Q}$ is tensor of rank two; $\mathbb{K}^u$ stands for $\mathbb{K}^{u(q)}$, $\mathbb{M}$ is $\mathbb{K}^{w(l)}$, $\mathbb{N}$ is $\mathbb{K}^{w(q)}$, and $\mathbb{L}$ is $\mathbb{K}^{w(c)}$, in which $\mathbb{K}$ denotes tensor of higher ranks. $\alpha$ is the optical modal coordinate, and $p_i$ and $q_{nm}$ are, respectively, the generalized coordinates of the membrane's in-plane radial and out-of-plane motions. Further, $\mathbb{K}^{u(q)}$ returns to the EM's in-plane quadratic stiffness, and $\mathbb{K}^{w(l)}$, $\mathbb{K}^{w(q)}$, and $\mathbb{K}^{w(c)}$ denote its out-of-plane linear, quadratic, and cubic stiffness components. Equations (ES22), (ES23), and (ES25) form a set of nonlinearly coupled equations describing the dynamics of the elastomer cavity-optomechanic device.

Figure S1 depicts the membrane's frequency-displacement behaviour as the acoustic forcing frequency is swept in the vicinity of the fundamental out-of-plane mode's natural frequency of the elastomer. Increasing the forcing frequency from lower values enhances the membrane's motion amplitude until it reaches point A, experiencing a jump down to the low-amplitude response branch at point B. Further increasing the forcing frequency, the membrane's displacement decreases gradually along this branch. By backward sweeping the forcing frequency, the periodic orbit's amplitude grows along the right low-amplitude branch until it reaches point C. At this point, the membrane's motion observes an abrupt jump up to the large-amplitude response branch landing on point D, as the sweeping parameter is decreased further. The response amplitude decreases further as the external excitation frequency is reduced to the lower frequencies. This scenario leads the frequency-response curve's branches bend to the right, introducing hardening behaviour, reflecting mid-plane stretching effect that arises from clamped boundary conditions. The steady-state dynamics of the elastomer membrane in the absence of the optical cavity is shown in Figure S2a, for four values of the external forcing amplitude, 0.25, 1, 1.5, and 2. As observed, the frequency-displacement behaviour is linear and made of only a continuous branch for small forcing amplitude, $F_{ac} = 0.25$. By driving the membrane structure with a relatively harder external excitation, the frequency-response branches experience jump phenomenon at points A and C through cyclic-fold bifurcation. The influence of Poisson's ratio on the membrane's steady-state dynamics is demonstrated in Figure S2b. It is seen that deviating from incompressibility assumption, $\nu = 0.5$, enhances the hardening degree of the membrane's motion. However, it is saturated as Poisson's ratio reaches a minimum threshold, which is approximately $\nu = 0.3$.

Figure S3 illustrates the dynamic response of the elastomer membrane-cavity optomechanic system in the absence of the external acoustic wave. Numerical integration is schemed to capture the solution to the nonlinear optomechanical equations. For a low optical input flux, $|s_{in}| = 1 \times 10^8 \text{Hz}^{-0.5}$, which is lower than a threshold, the optical modal response is stationary so that its frequency spectrum contains only a single bias component, Figures S3a and S3b. However, increasing the input photon flux up to $|s_{in}| = 8 \times 10^{10} \text{Hz}^{-0.5}$, triggers the interaction between the optical and mechanical sides of the system, generating a soliton frequency comb in the optical response frequency spectrum, Figure S3d. Figure S3c illustrates the corresponding time-domain response, indicating the emergence of equidistant combs in the temporal domain. Here, the detuning between the laser pump frequency ($\omega_L$) and the optical resonance frequency at rest ($\omega_0$) is set to be $\Delta = \omega_L - \omega_0 = -100$ (non-dimensional). It is worth noting that the generated frequency spectrum is stable having a solidarity profile as the optical pump frequency is varied within a detuning range from $-10000$ to $10000$. Moreover, it is important to mention that the system's free spectral ranging is locked at the membrane's first out-of-plane mode, $\text{FSR} = \omega_1$. On the other hand, the system exhibits a time-spacing equal to the period of the first mechanical mode, $t_{sp} = T_1 = \frac{2\pi}{\omega_1}$. Further amplifying the input light, $|s_{in}| = 4 \times 10^{11} \text{Hz}^{-0.5}$, drives the optomechanic system into richer soliton frequency combs extending its teeth over a broader dimensionless frequency ranging from -15000 to 1500, Figures S3e and S3f. For this case, the phase-space diagram of the mechanical side is plotted in Figure S3g, showing a periodic orbit that gradually contracts over time due to the influence of viscous damping embedded in the system's dynamics.

The optical power can be intensified via employing an external excitation whose frequency is set to drive the mechanical mode directly. This scenario is examined for the non-resonant case where the laser pump frequency is far below from the optical resonance frequency, although the input photon flux is above its required threshold for the resonant case, for instance: $\Delta = \omega_L - \omega_0 = -10^7$, and $|s_{in}| = 2 \times 10^{11} \text{Hz}^{-0.5}$. Figures S4(a) and S4(b) show the time- and the frequency-domain response of the optical modal coordinate

in the absence of the external forcing signal. By adjusting the acoustic wave frequency to match the natural frequency of the membrane's first out-of-plane mode, $\Omega = \omega_1$, we can generate a dual-state soliton frequency comb in the optical response, while the mechanical side displays quasi-periodic motion, Figures S4c and S4d. Here, the non-dimensional forcing amplitude is set to be $F_{ac} = 0.1$. Figure S4e presents the optical frequency spectrum, depicting a two-state soliton frequency comb having a spacing corresponding to the acoustic wave frequency within the range of $-5000$ to $5000$. Elevating the forcing amplitude to $F_{ac} = 0.2$ results in a similar scenario but with a broader frequency spectrum, Figures S5. Moreover, as seen in Figure S5(d), the membrane's phase-space diagram introduces a quasi-periodic orbit that is damped over time. Making comparison between Figures S3 and S4, we can conclude that the optical side generates a considerably broader frequency spectrum with reduced power when the cavity optomechanical system is externally driven, while maintaining a separation between the laser pump frequency and the optical resonance frequency.

This aligns with our experimental findings regarding the generation of multiple soliton frequency combs when the elastomer membrane-cavity system is externally excited using an acoustic waveform.

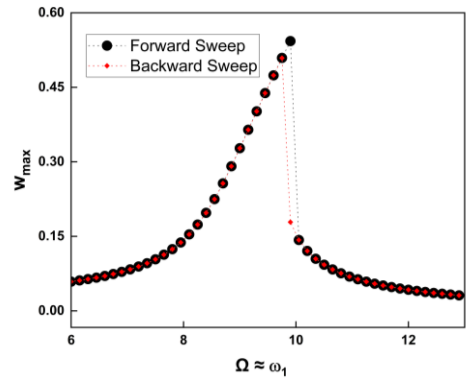

**Supplementary Figure 1**: *Frequency-response curve of the elastomer membrane-based microresonator obtained from numerical simulation. The membrane is driven by an externally applied harmonic wave in the form of $F_{ac}\,cos(\Omega t)$ with the force amplitude of $F_{ac} = 0.1$. The forcing frequency is varied near the first out-of-plane mode's natural frequency to drive the membrane's primary resonance, $\Omega \approx \omega_1$.*

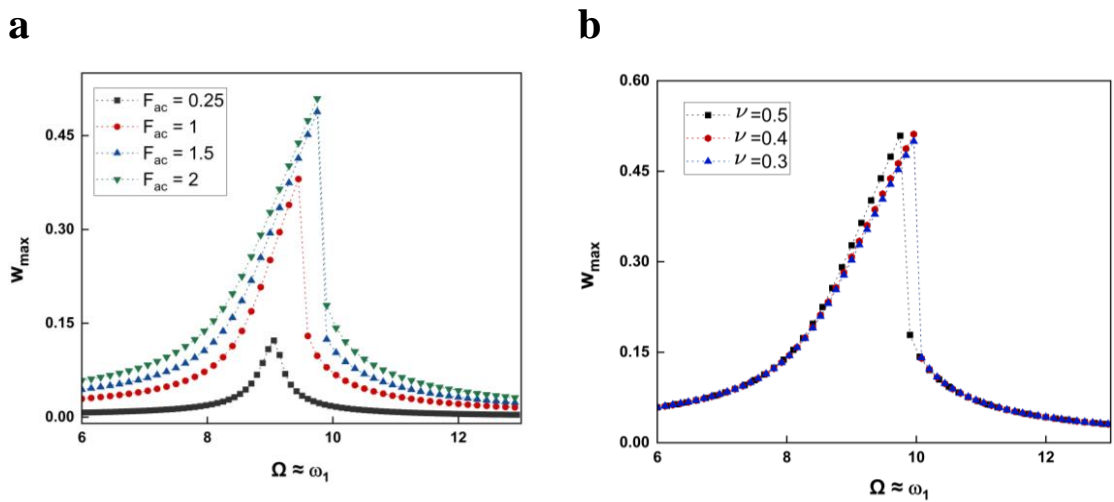

**Supplementary Figure 2**: *The simulated steady-state dynamic response of the EM microresonator to an external harmonic excitation in the vicinity of its primary resonance. **a,** The EM's frequency-displacement curves simulated for four values of the forcing amplitude, $F_{ac} = 0.25, 1, 1.5, 2$. The membrane exhibits hardening behaviour, experiencing two cyclic-fold bifurcations on the large- and small-amplitude solution branches, for large values of the forcing amplitude. **b,** The influence of Poisson's ratio on the loci of the periodic orbit's amplitude, for $F_{ac} = 1$.*

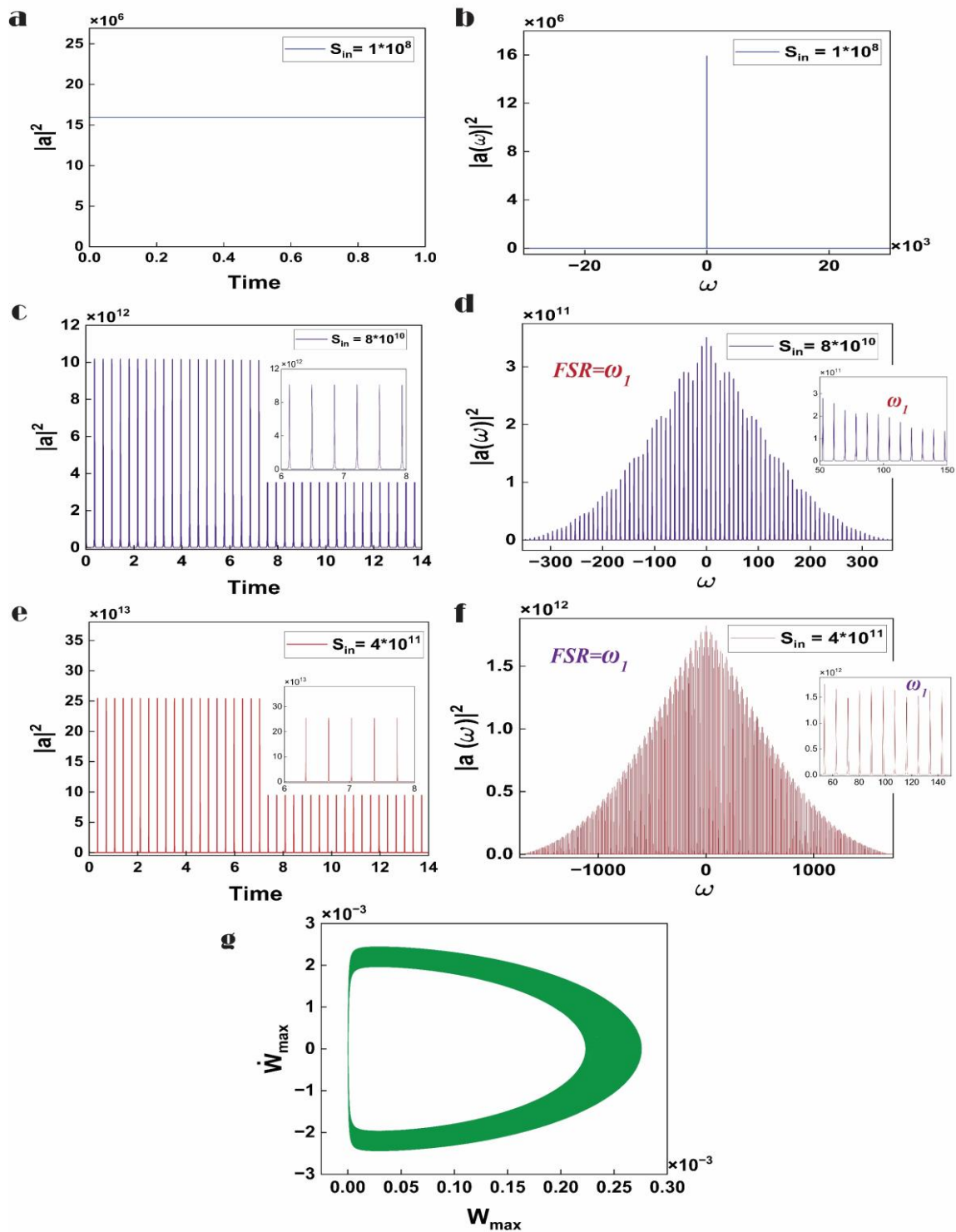

**Supplementary Figure 3**: *The simulated dynamics of the EM-cavity optomechanical system in the absence of the external acoustic waveform, for the optical detuning frequency of* $\Delta = -10^2$. ***a*** *and* ***b,*** *respectively, depict the time-domain response and the corresponding frequency spectrum of the input photon flux, for* $|s_{in}| = 1 \times 10^8 \text{ Hz}^{-0.5}$. *Increasing the input optical power to* $|s_{in}| = 8 \times 10^{10} \text{ Hz}^{-0.5}$ *intensifies the intracavity*

*Kerr nonlinearity, driving the system into an extreme nonlinear regime generating soliton frequency combs.* **c** *and* **d,** *illustrate the generation of combs in the time- and frequency-domain, respectively, having the spacing of $t_{sp} = XX$ and $FSR = \omega_1$. Further enhancing the input optical flux to $|s_{in}| = 4 \times 10^{11} \text{ Hz}^{-0.5}$, the bandwidth of the soliton frequency combs expands, covering a dimensionless frequency range from -1000 to 1000, Figures 3**e** and 3**f**. **g,** The EM's phase-space diagram corresponding to Figure 3**e**.*

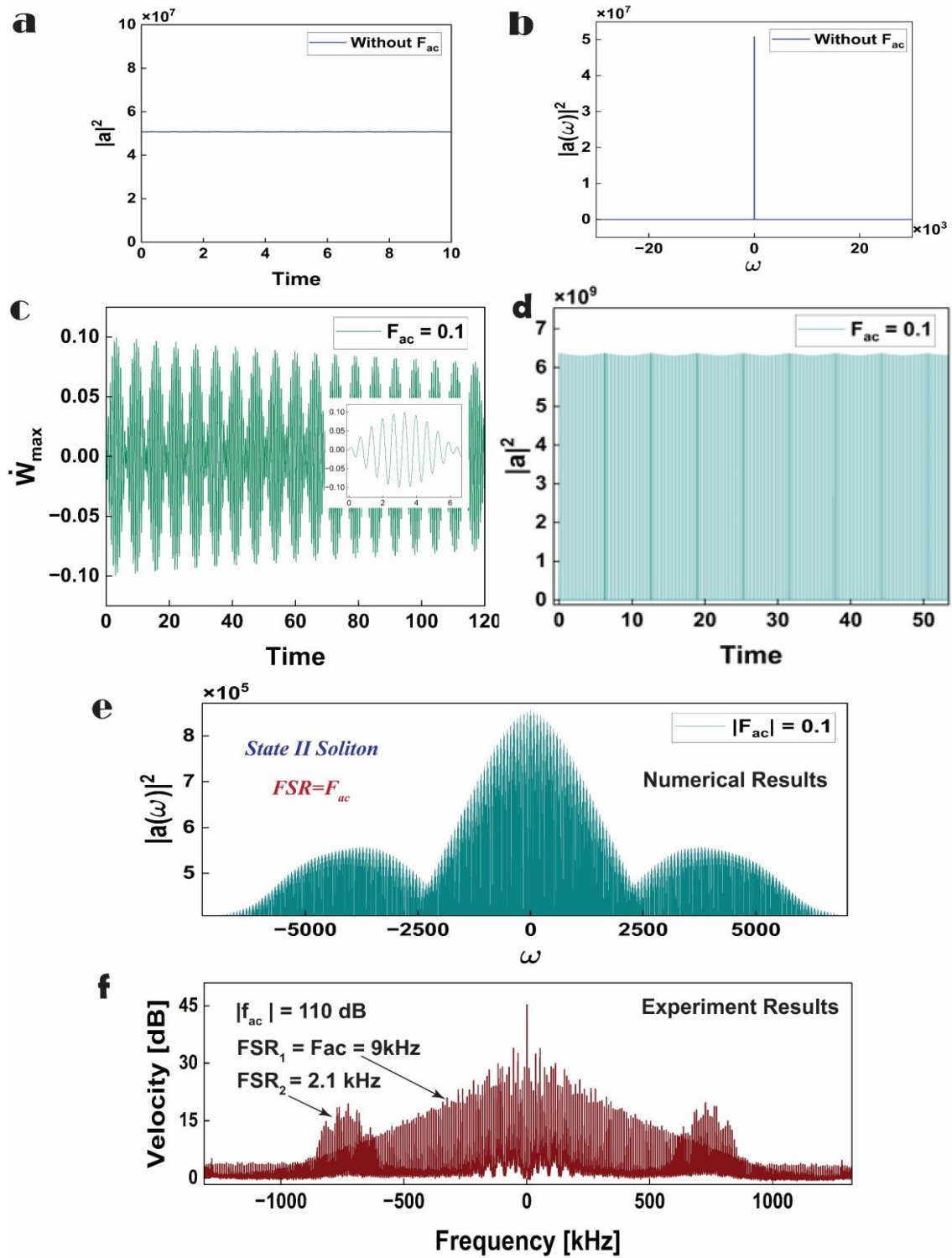

**Supplementary Figure 4**: *The simulated dynamic response of the EM-cavity optomechanical device, for the optical detuning frequency of $\Delta = -10^7$ (non-resonant case). **a** and **b**, respectively, demonstrate the time history and the corresponding frequency content of the optical modal coordinate in the absence of the external*

waveform, for $|s_{in}| = 2 \times 10^{11} \, \text{Hz}^{-0.5}$. **c** and **d**, respectively, show the time-domain evolution of the system's mechanical displacement and optical response when the EM is externally driven using a harmonic signal with the amplitude of $F_{ac} = 0.1$ and frequency of $\Omega = \omega_1$. **e,** The frequency spectrum of the optical response indicating a dual-state soliton frequency comb spanning a dimensionless frequency range from $-5000$ to 5000. **f,** the experimental generated soliton FCs for the case of CW (1mW, 680 nm and 726.65 kHz) and $|F_{ac}|$=110 dB, exciting the second mechanical mode.

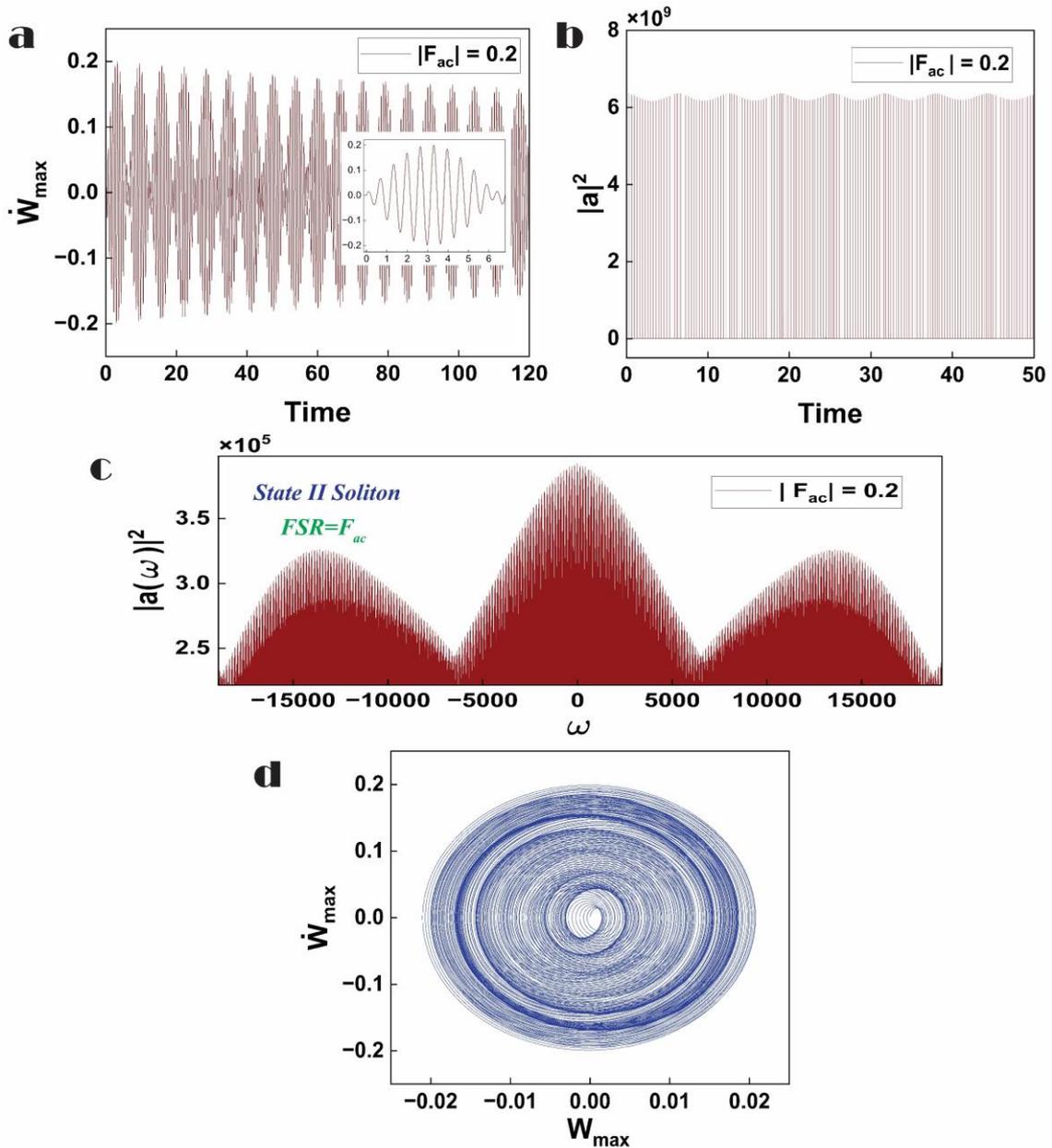

**Supplementary Figure 5**: *The simulated dynamic response of the EM-cavity optomechanical device, for the optical detuning frequency of $\Delta = -10^7$ (non-resonant case), and the injected photon flux of $|s_{in}| = 2 \times 10^{11}$ $Hz^{-0.5}$. **a** and **b**, respectively, show the time-domain evolution of the EM's displacement and the system's optical response in the presence of the external acoustic wave with the amplitude of $F_{ac} = 0.2$ and frequency of $\Omega = \omega_1$. **c,** Depicts the frequency spectrum of the optical response indicating a dual-state soliton frequency comb spanning a dimensionless frequency range from $-15000$ to 15000. **d,** The mechanical side's trajectory plotted in a 2D phase-space capturing a damped quasi-periodic motion.*